# Potential Antimicrobial Activity of Marine Sponge *Neopetrosia exigua*

Ibrahim Majali[1], Haitham N. Qaralleh[1*], Syed Z. Idid[2], Shahbudin Saad[3], Deny Susanti[2], Osama Y. Althunibat[1]

[1]Faculty of Pharmacy, Al-Zaytoonah Private University, Amman 11733, Jordan.
[2]Department of Biomedical Science, Faculty of Science, International Islamic University Malaysia, Kuantan, Pahang, Malaysia
[3]Institute of Oceanography and Maritime Studies, Faculty of Science, International Islamic University Malaysia, Kuantan, Pahang, Malaysia

*corresponding author: haithym2006@yahoo.com

**Abstract:** *Neopetrosia exigua* has received great attention in natural product chemistry. The diversity of *N. exigua* constituents has been demonstrated by the continued discovery of novel bioactive metabolites such as antimicrobial metabolites. In this study, in order to localise the active component of *N. exigua* biomass according to the polarity, a sequential gradient partition with different solvents (*n*-hexane, carbon tetrachloride, dichloromethane, *n*-butanol, and water) was performed to obtain fractions containing metabolites distributed according to their polarity. The antimicrobial activities of *N. exigua* fractions were then evaluated using disc diffusion and microdilution methods (influence on the growth curve, Minimum Inhibitory Concentration (MIC) and Minimum Bactericidal Concentration (MBC)). The results showed that the active metabolites were present in *n*-hexane, $CH_2Cl_2$, *n*-BuOH, and water fractions. *n*-hexane, $CH_2Cl_2$, and *n*-BuOH fractions were the most effective fractions. Among microbes tested, *Staphylococcus aureus* was the most susceptible microbe evaluated. The obtained results are considered sufficient for further study to isolate the compounds represent the antimicrobial activity.

## INTRODUCTION

The development of drug resistance in human pathogens against commonly used antibiotics and the decline in antibiotic discovery have necessitated a search for new antimicrobial substances from other sources including natural sources from any terrestrial or marine source (Sneader, 2005; Sanglard et al., 2009). Marine organisms are known to produce different compounds to protect themselves against the variety of their own pathogens and therefore can be considered as a potential source of many classes of antimicrobial substances. More than 18,000 natural products, used for different purpose, have been mainly isolated from marine sponges, jellyfish, sea anemones, corals, bryozoans, molluscs, echinoderms, tunicates and crustaceans. However, sponges, which are the most primitive invertebrates, are considered as the major rich phyla with novel bioactive compounds (Blunt et al., 2007).

There is no doubt that sponges contain antimicrobial active metabolites localised in their body. More than 800 antibiotic metabolites have been isolated from marine sponges (Torres et al., 2002), and many new metabolites are being discovered every year. Although none of the discovered metabolites has been developed as an antibacterial drug, many are currently under clinical investigation. Psammaplin A, which was isolated from the sponge Psammaplysilla (Jimenez and Crews, 1991), is the first antibacterial metabolite from sponges that is going to be commercialised soon (Laport et al., 2009).

In fact, sponges that produce antimicrobial metabolites always host symbiotic bacteria (Kobayashi and Ishibashi, 1993; Thakur and Anil, 2000; Thomas et al., 2010). In some cases, about 50-60% of the sponge biomass is represented by associated microbes (Vacelet and Donadey, 1977; Willenz and Hartman, 1989). But, there is no proof whether the bioactive metabolites are produced by the sponge-associated microorganisms or by the sponge itself (Unson and Faulkner, 1993; Faulkner et al., 1994; Unson et al., 1994; Brantley et al., 1995; Bewley et al., 1996; Lin et al., 2001). In other words, sponge metabolites could be synthesised by the sponge itself (Garson et al., 1992; Pomponi and Willoughby, 1994) or it is obtained from other sources. The sources are thought to be limited to the symbiotic microbes or to the free living microbes in the marine environment (Lindquist et al., 2005).

*Neopetrosia exigua* Kirkpatrick belongs to the phylum Porifera, the class Demospongia, of the order Haplosclerida, and family Petrosiidae. *N. exigua* is a tropical marine sponge, dominant in tropical Australasia, Western South Pacific and South East Asia. It is brownish in colour and it has encrusting base with erect lamellae (Hooper et al., 2000; Cheng et al., 2008). *N. exigua* has received great attention in natural product chemistry. The diversity of *N. exigua* constituents has been demonstrated by the continued discovery of novel bioactive metabolites. Since 1980's, more than 24 metabolites have been isolated from *N. exigua* (Table 1). These metabolites have been reported to possess variable bio-activities including vasodilation, cytotoxicity, antibacteria, and other activities. The metabolites isolated from *N. exigua* showed variation in their chemical structures such as alkaloids, quinones, and sterols (Roll et al., 1983; Nakagawa and Endo, 1984; Williams et al., 1998; Orabi et al., 2002; Iwagawa et al., 2000; Williams et al., 2002; Cerqueira et al., 2003).

On the other hand, most of the isolation strategies reported from *N. exigua* were focused on the isolation of metabolites from the whole sponge biomass. Only two studies isolated novel antimicrobial metabolites from *N. exigua*-associated microbes. One of the isolated metabolites is xestodecalactone, which was isolated from the fungus *Penicillium cf. montanense* and this metabolite showed antifungal activity (Edrada et al., 2002). Other isolated metabolite is aspergiones, which were isolated from *Aspergillus versicolor*. Aspergiones were reported to possess antibacterial activity (Lin et al., 2001).

In our previous preliminary screening (Qaralleh et al., 2010), *N. exigua* extracts showed promising antimicrobial activity. After that, it was hypothesised that the antimicrobial metabolites of *N. exigua* could be variable according to their polarity since different classes of chemical structures were isolated previously from this species.

**MATERIAL AND METHODS**
**Sampling and Identification**
The marine sponge *N. exigua* was collected from Langkawi Island - Malaysia (6°12′47.01″N 99°44′39.32″E) in 2009 at depths of 1-2 m. The collected sponge was transported to the lab in an ice box containing dry ice. *N. exigua* materials were cut into small pieces and immediately frozen and maintained at -20°C prior to extraction.
The identification of *N. exigua* was carried out by Mr. Lim Swee-Cheng (Tropical Marine Science Institute, National University of Singapore, Singapore) based on skeletal slides and dissociated spicule mounts.

Table 1: Bioactive metabolites isolated from *N. exigua*

| Metabolite | Source | Class | Bio-activity | References |
|---|---|---|---|---|
| (+)-Xestospongin A-D | Biomass | Alkaloids | Vasodilation | Nakagawa and Endo (1984) |
| (-)-Araguspongine K and L | Biomass | Alkaloids | Vasodilation | Orabi et al., (2002) |
| Xestosin A | Biomass | Alkaloids | Not detected | Iwagawa et al., (2000) |
| Motuporamine A- I | Biomass | Alkaloids | Cytotoxicity | Williams et al., (1998); Williams et al., (2002) |
| 9'-Epi-3β,3'β-dimethylxestospongin | Biomass | Alkaloids | Not detected | Li et al., (2011) |
| Exiguamine A | Biomass | Alkaloid | Inhibition of indoleamine-2,3-dioxygenase | Brastianos et al., (2006) |
| Halenaquinone | Biomass | Quinones | Antibacterial | Roll et al., (1983) |
| Exiguaquinol (**2**) | Biomass | Quinones | Inhibit *Helicobacter pylori* glutamate racemase (Mur I) | De Almeida Leone et al., (2008) |
| Clionasterol | Biomass | Sterols | Inhibit human complement system | Cerqueira et al., (2003) |
| 5α,8α-Epidioxy-24α-ethylcholest-6-en-βb-ol | Biomass | Sterols | Inhibit human complement system | Cerqueira et al., (2003) |

Metabolites isolated from *Xestospongia exigua* were included in this table since *X. exigua* was re-classified as *Neopetrosia exigua*.

**Extraction**
The frozen *N. exigua* materials were thawed, washed, and then freeze dried in order to remove the excess amount of water. The freeze dried materials were homogenised and extracted three times by absolute methanol (MeOH) for 24 h each. The supernatants were removed, centrifuged at 10000×g, filtred, and collected. The MeOH extracts were combined and the solvent was removed by rotary evaporation. The MeOH extract of *N. exigua* biomass was kept in a deep freezer at -20°C until it was used.

**Liquid-liquid fractionation**
A sequential gradient partition with different solvents was performed to obtain fractions containing compounds distributed according to their polarity. Briefly, MeOH extract of *N. exigua* biomass was partitioned successively with *n*-hexane, carbon tetrachloride ($CCl_4$), dichloromethane ($CH_2Cl_2$), *n*-butanol (*n*-BuOH), and water (Riguera, 1997). After evaporating the solvents using a rotary evaporator, the crude extracts of each solvent were labelled and stored (-20°C) until it was used.

**Antimicrobial assays**
**Samples preparation**
A sample of 100 mg from each extract was dissolved in 1 mL methanol-deionised water (4:1). The extract was then clarified by filtration through sterile syringe filter with 0.2-0.45 μm pore. Finally, the filtered extract was stored as aliquots until it was used.

**Microbial strains**
Six reference strains of human pathogens were used including two Gram-positive strains (*Staphylococcus aureus* ATCC25923 and *Bacillus cereus* ATCC11778), two Gram-negative strains (*Pseudomonas aeruginosa* ATCC27853 and *Escherichia coli* ATCC35218), and two yeast strains (*Candida albicans* ATCC10231 and *Cryptococcus neoformans* ATCC90112).

**Disc diffusion method**
Antimicrobial activities were tested using disc diffusion method. Briefly, inoculum containing $10^7$ CFU/mL was spread on Mueller-Hinton agar (MHA) plates for bacteria and $10^4$ CFU/mL was spread on potato dextrose agar (PDA) for yeast. Using sterile forceps, the sterile filter papers (6 mm diameter) containing different concentrations of the extracts, negative control or positive control were laid down on the surface of inoculated agar plate. The plates were incubated at 37°C for 24 h for the bacteria and at room temperature (18-20°C) for 24-48 h for the yeast strains. Each sample was tested in triplicate and the zone of inhibition was measured as millimetre diameter.

**Influence of *N. exigua* fractions on the microbial growth curves of Gram-positive bacteria and yeast strains:**
The influence on the growth curves was determined by using microdilution method. Three fold dilutions were prepared from the fractions tested in this study. This was performed using 96-well plate and by mixing 30 μL of the fractions (stock concentration of 10 mg/mL) with 270 μL broth media in the first well of each column to produce 1000 μg/mL of final concentration. After mixing by pipetting, 100 μL from the well containing 1000 μg/mL extract (first well of each column) was carried over into the next well in the same column that contained 200 μL broth media. This step was repeated to the end of each column of the plate. For each fraction, three columns were treated with the assigned microbe while the fourth column was used as a blank. In each plate, 8-wells filled with the broth media and infected with the assigned microbes were used as control.

The final microbes inoculum was adjusted to $10^7$ CFU/mL for bacteria and $10^4$ for the yeasts strains then 10 μL was added to each test well. The microplates were incubated for 12 h at 37°C for bacteria and at room temperature for yeasts inside the microplate reader. The turbidity of the cultures was taken every 1 h for 12 h as indicator of the bacterial growth. The turbidity in each well was measured at 600 nm for the bacteria and at 490 nm for the yeast strains. The turbidity value of blank well was subtracted from the turbidity value obtained from the treated well for each concentration. The growth curves of the tested microbes in absence (normal growth) or presence of different concentrations of the fractions were depicted using Microsoft Excel 2010.

**Determination of Minimum Inhibitory Concentration (MIC)**
Further incubation of the cultures (96-well plates) that were used in the growth curves experiments was performed for 12 h (total 24 h) and the turbidity was measured as before. The lowest concentration of fraction needed to inhibit the visible growth of a test microorganism after 24 h was considered as the MIC.

**Determination of Minimum Bactericidal Concentration (MBC):**
Minimal bactericidal concentration (MBC) was determined by transferring and spreading the treated culture broth of the wells containing the concentrations of equal to and higher than the MIC on agar plates. The lowest concentration of the fraction required to completely destroy test microorganism (no growth on the agar plate) after incubation at 37 °C for 24 h (bacteria) or at room temperature for 48 h (yeasts) was reported as MBC.

**RESULTS**
*N. exigua* extract was examined at concentrations of 0.4, 1, 2, and 2.5 mg/disc against the Gram-positive *B. cereus* and *S. aureus*, Gram-negative *E. coli* and *P. aeruginosa*, and the yeast strains *C. albicans* and *C. neoformans*. Generally, the results showed that the MeOH extract of *N. exigua* biomass exhibited dose-dependent activity as the inhibition activity of the extract increased by increasing the concentration (Figure 1).
Gram-negative bacteria appeared to be insensitive to the extract as the maximum inhibition zone observed was 9.5 mm against *E. coli*. Meanwhile, *P. aeruginosa* was highly resistant to the extract since no inhibition zone was observed even at the highest concentration tested. On the contrary, Gram-positive bacteria were in general the most sensitive strains (Figure 2). At all concentrations tested, the extract showed remarkable inhibition zones. The range of inhibition zones observed was between 12.8 and 24.5 mm. Indeed, *S. aureus* (Figure 2D) was less sensitive than *B. cereus* (Figure 2A-C), which gave the highest inhibition zone observed (24.5 mm). For the yeast strains tested, both *C. albicans* and *C. neoformans* appeared to be sensitive to *N. exigua* biomass extract (inhibition zones ranging from 8 to 22.8 mm). However, *C. neoformans* (inhibition zones ranging from 10 to 22.8 mm) was more sensitive than *C. albicans* (inhibition zones ranging from 8 to 19.8 mm).

**Antimicrobial activity of *N. exigua* fractions using disc diffusion method**

A sequential gradient partition with solvents of different polarities was carried out using *n*-hexane, $CCl_4$, $CH_2Cl_2$, *n*-BuOH, and water. The fractions obtained contain compounds distributed according to their polarity.

Regardless of the concentrations tested, all fractions except the $CCl_4$ fraction showed inhibition zones against *S. aureus* and *B. cereus* (Figures 3A and B; Figure 4B). Similar to *N. exigua* biomass extract, all active fractions tested showed dose-dependent inhibition activities as the size of inhibition zone increased by increasing the concentration. The highest antimicrobial activity observed was for $CH_2Cl_2$ fraction followed by the fractions of *n*-BuOH, *n*-hexane, and water. The absence of the inhibition zones for the water fraction at the lowest concentration tested point to it as having the weakest activity (Figure 4G). For the *n*-hexane fraction, the highest inhibition zone observed was 20.7 mm against *S. aureus* (Figure 4I) while the lowest inhibition zone observed was 0 mm against *B. cereus*. *S. aureus* (range of inhibition zones observed was 14.3 to 20.7 mm) appeared to be more susceptible to *n*-hexane fraction than *B. cereus* (range of inhibition zones observed was 0 to 15.8 mm).

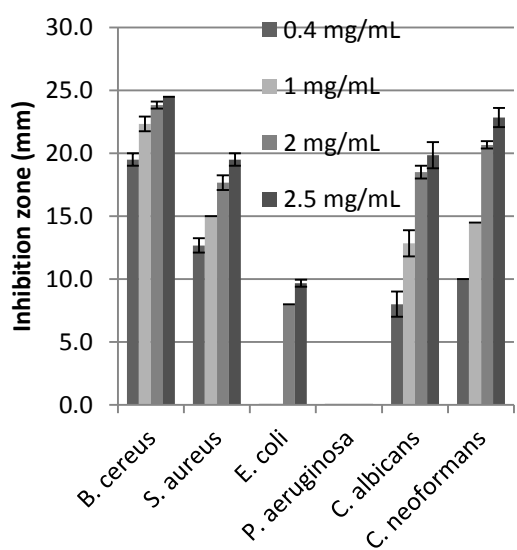

**Figure 1:** Antimicrobial activities of different concentrations of *N. exigua* extract using disc diffusion method

Both $CH_2Cl_2$ and *n*-BuOH fractions showed remarkable inhibition zones even at the lowest concentration tested (Figure 3A and B). Both *B. cereus* and *S. aureus* were sensitive to the $CH_2Cl_2$ and *n*-BuOH fractions but the inhibition activity against *B. cereus* was higher. $CH_2Cl_2$ fraction possessed stronger antimicrobial activity than *n*-BuOH fraction. The range of inhibition zones observed for the $CH_2Cl_2$ fraction was 23.8 to 28.7 mm and 21.3 to 26.7 mm against *B. cereus* (Figures 4C and D) and *S. aureus*, respectively. On the contrary, the ranges of inhibition zones observed for the *n*-BuOH fraction against *B. cereus* (Figure 4E and F) and *S. aureus* were 17.8 to 23.7 mm and 14.5 to 22 mm, respectively.

Two yeast strains namely *C. albicans* and *C. neoformans* were used to evaluate the antifungal activities of *N. exigua* fractions (Figure 5). Generally, while *n*-hexane and $CCl_4$ fractions have no antifungal activities, fractions of $CH_2Cl_2$, *n*-BuOH and water showed variable antifungal activities. Dose-dependent activity was also observed for the antifungal activity of the active fractions. Although the lowest concentrations of the water fraction tested showed no or negligible inhibition zones (0 to 8.3 mm), the highest inhibition zone observed from the water fraction was 12.7 mm at 2.5 mg/disc. The presence of inhibition zone of 21.3 mm and 18.3 mm indicated strong antifungal activities of the $CH_2Cl_2$ and *n*-BuOH fractions. It is observed that *C. neoformans* was more susceptible than *C. albicans*. The inhibition zone observed against *C. neoformans* was in the range of 10.3 to 21.3 mm and the range was 8.7 to 19.5 mm against *C. albicans*.

**Influence of *n*-hexane, $CH_2Cl_2$, and *n*-BuOH fractions on the microbial growth curves**

Fractions that showed maximum antimicrobial activity using disc diffusion method were investigated in this test. The fractions were tested at different concentrations of equal to 1000, 333, 111, 37, 12.3, and 4.1 µg/mL. The turbidity of the cultures was taken every 1 h for 12 h as indicator of microbial growth. The growth curves of the tested microbes in absence or presence of the fractions were depicted in Figures 6 and 7. The normal growth curves of all tested microbes were depicted in the same figures for comparison.

All microbes tested showed similar lag phase period. After 3 h of incubation, the growth showed steady increase, thus the starting point of the exponential phase was noted for all microbes. Although, *S. aureus, C. albicans*, and *C. neoformans* tended to enter the stationary phase after 7 to 9 h, *B. cereus* showed no stationary phase during the experiment period (12 h). Of all concentrations tested, the influence of the fractions on the microbes' growth curves can be divided into two categories. First, the concentrations that completely inhibited the microbes' visible growth. Second, the concentrations that showed dose-dependent inhibition activities. But the effect of the fractions on the lag phase was not critical compared to the effects of the fractions on the exponential and stationary phases. Therefore, detailed descriptions were given for the influence of each fraction according to these points.

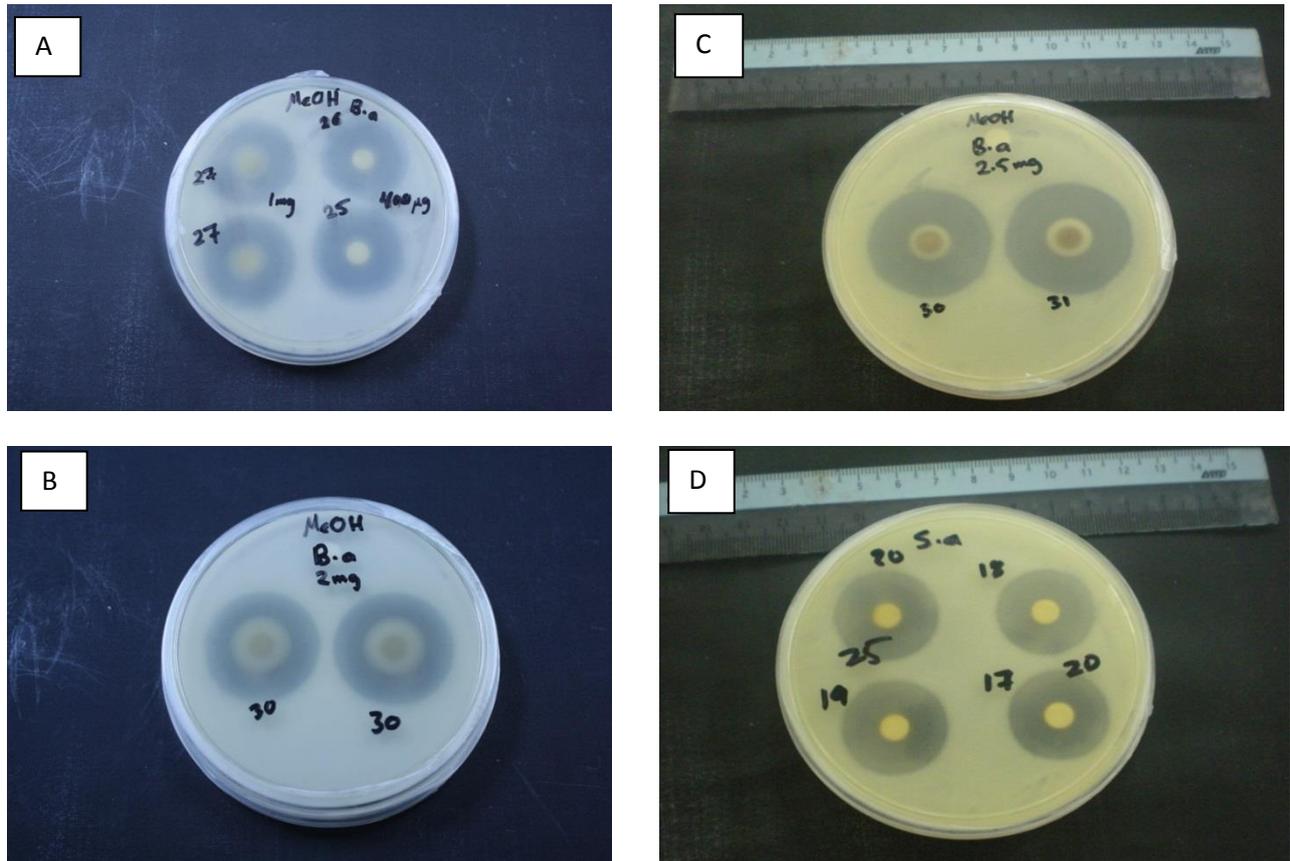

**Figure 2:** Inhibition zones observed from different concentrations of *N. exigua* biomass extract using $10^7$ CFU of *B. cereus* and *S. aureus*. A: *B. cereus* treated with 0.4 and 1 mg/disc, B: *B. cereus* treated with 2 mg/disc, C: *B. cereus* treated with 2.5 mg/disc, D: *S. aureus* treated with 2 and 2.5 mg/disc.

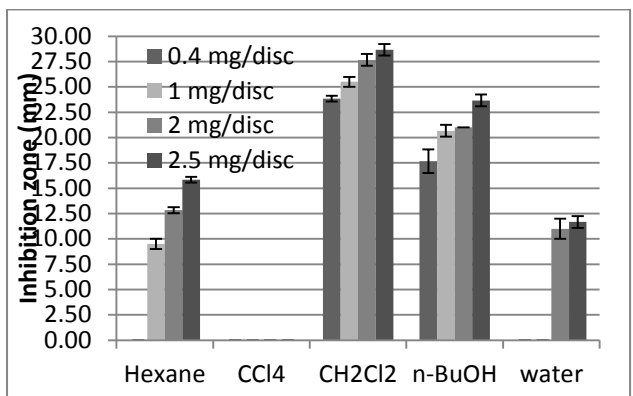
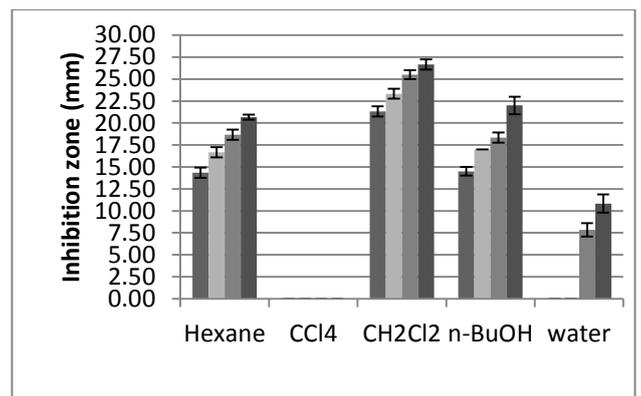

**Figure 3:** Antimicrobial activities of different concentrations of *N. exigua* fractions (*n*-hexane, CCl₄, CH₂Cl₂, *n*-BuOH, and water) using disc diffusion method. The fractions were tested at different concentrations (0.4, 1, 2, 2.5 mg/disc) against *B. cereus* (A) and *S. aureus* (B). Data represent mean diameter of triplicate of zone of inhibition including the diameter of the disc (6 mm) and expressed as mm±SD.

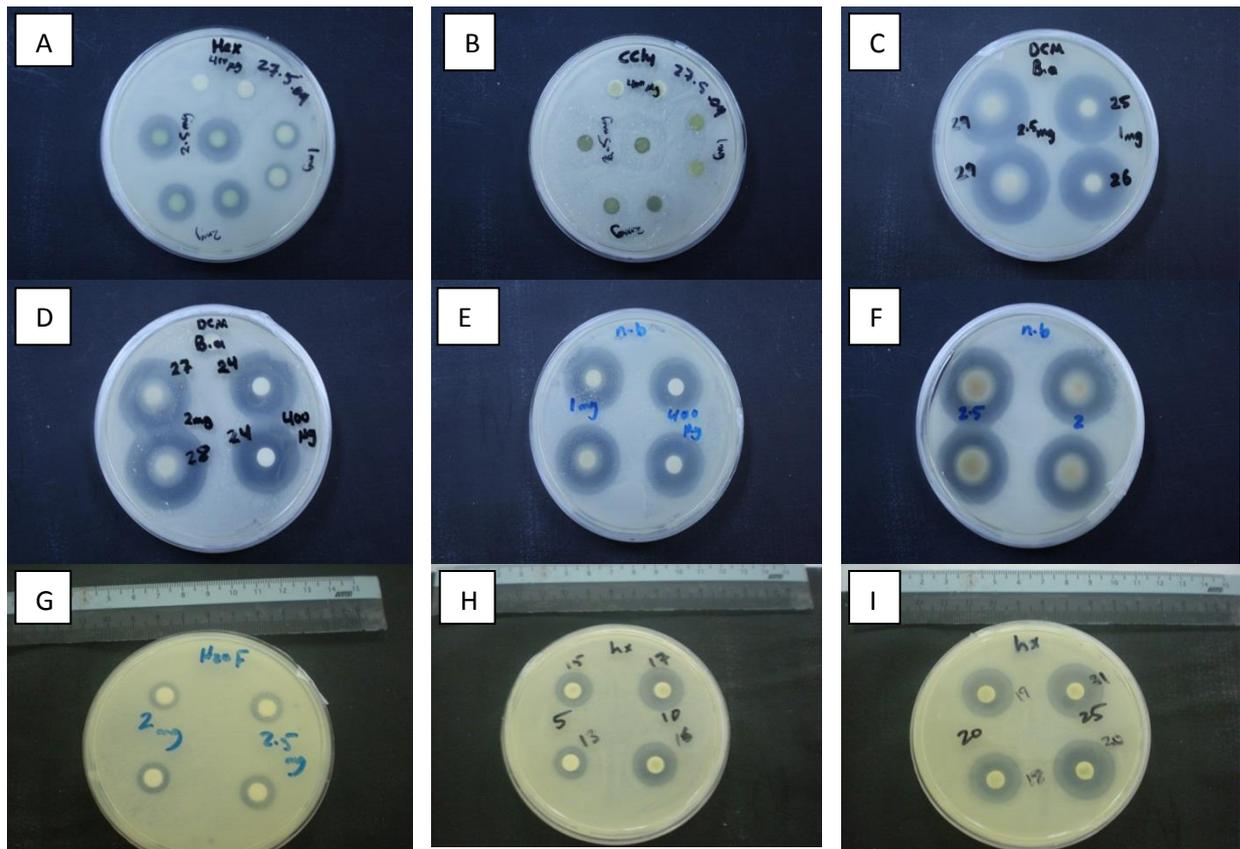

**Figure 4:** Inhibition zones observed from different concentrations of *N. exigua* fractions (*n*-hexane, $CCl_4$, $CH_2Cl_2$, *n*-BuOH, and water) using $10^7$ CFU of *B. cereus* and *S. aureus*. A: *B. cereus* treated with *n*-hexane fraction (0.4 to 2.5 mg/disc), B: *B. cereus* treated with $CCl_4$ fraction (0.4 to 2.5 mg/disc), C: *B. cereus* treated with $CH_2Cl_2$ fraction (1 and 2.5 mg/disc), D: *B. cereus* treated with $CH_2Cl_2$ fraction (0.4 and 2 mg/disc), E: *B. cereus* treated with *n*-BuOH fraction (0.4 and 1 mg/disc), F: *B. cereus* treated with *n*-BuOH fraction (2 and 2.5 mg/disc), G: *B. cereus* treated with water fraction (2 and 2.5 mg/disc), H: *S. aureus* treated with *n*-hexane fraction (0.4 and 1 mg/disc), I: *S. aureus* treated with *n*-hexane fraction (2 and 2.5 mg/disc).

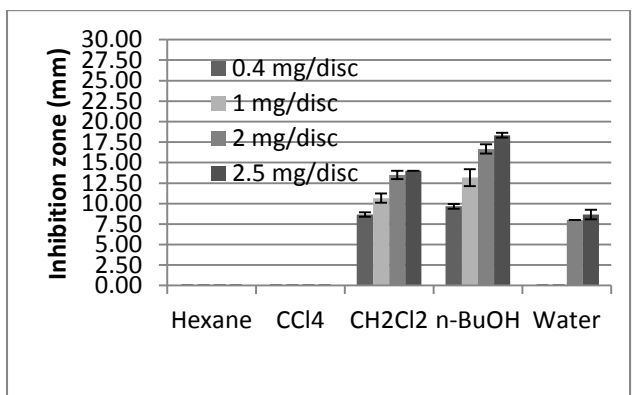 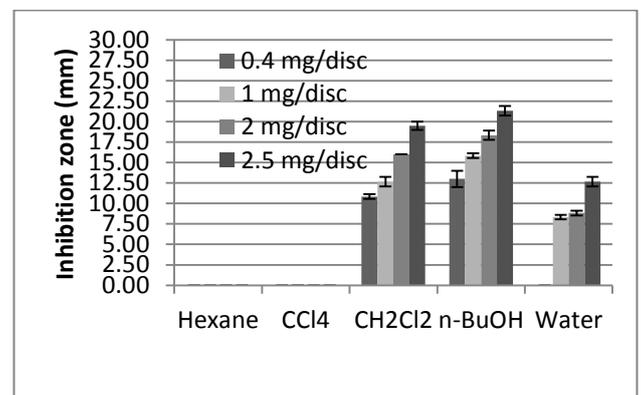

**Figure 5**: Antifungal activities of different concentrations of *N. exigua* fractions (*n*-hexane, $CCl_4$, $CH_2Cl_2$, *n*-BuOH, and water) using disc diffusion method. The fractions were tested at different concentrations (0.4, 1, 2, 2.5 mg/disc) against *C. albicans* (A) and *C. neoformans* (B). Data represent mean diameter of triplicate of zone of inhibition including the diameter of the disc (6 mm) and expressed as mm±SD.

**Influence on the growth curves of *B. cereus* and *S. aureus*:**

The results showed that the most effective concentration of *n*-hexane fraction was 1000 µg/mL, which completely inhibited the visible growth of *B. cereus* and *S. aureus*. The concentration of 333 µg/mL completely terminated the exponential phase of *B. cereus* in 1 hour while the concentration of 111 µg/mL caused early termination of the *B. cereus* exponential phase in 3 h (Figure 6A). Normal growth patterns from *B. cereus* growth curves were detected for all concentrations below 111 µg/mL while the growth curve of *S. aureus* at concentration of 4.1 µg/mL appeared to be the closest growth curve to the normal growth curve over the experiment time period. For *S. aureus*, concentrations below 1000 µg/mL showed dose-dependent inhibition activities (Figure 6B). When compared to the normal growth curve, the concentration of 333 µg/mL led to delayed beginning of the exponential phase of *S. aureus* for about 4 h.

For $CH_2Cl_2$ fraction, the results showed that the concentrations that completely inhibited the visible growth of *B. cereus* were all concentrations of higher than 37 µg/mL (Figure 6C). Dose-dependent inhibition of *B. cereus* growth was detected for all concentrations below 111 µg/mL. The beginning of exponential phase of *B. cereus* was delayed for 1 hour when it was treated with 37 µg/mL while the stationary phase started after 6 h incubation. On the other hand, visible growth of *S. aureus* was completely inhibited at all concentrations higher than 12.3 µg/mL. The growth of *S. aureus* was also inhibited at 12.3 µg/mL but recovery in the growth, representing the beginning of exponential phase, was observed after 10 h. A 2 h delay in the exponential phase of *S. aureus* was noted when it was treated with 4.1 µg/mL $CH_2Cl_2$ fraction.

For *n*-BuOH fraction, all concentrations higher than 37 µg/mL completely inhibited the visible growth of *B. cereus* and *S. aureus*. Dose-dependent inhibition activity was detected for all concentrations below 111 µg/mL. When *B. cereus* was treated with 12.3 and 4.1 µg/mL, the growth decreased compared to the growth of normal cells (Figure 6E). However, the treated *B. cereus* showed normal beginning of exponential phase. Moreover, concentration of 37 µg/mL led to the delay in the beginning of exponential phase for about 3 h. Similar to $CH_2Cl_2$ fraction, a recovery in the growth was noted after 10 h incubation when *S. aureus* was treated with 37 µg/mL (Figure 6F). A delay in the beginning of the exponential phase for 2 h and 1 hour was observed when *S. aureus* was treated with 12.3 and 4.1 µg/mL, respectively.

**Influence on the growth curves of *C. albicans* and *C. neoformans***

For $CH_2Cl_2$ fraction, the complete inhibition of the visible growth of *C. albicans* and *C. neoformans* was observed at all concentrations higher than 37 µg/mL. A sign of growth recovery, after 12 h incubation, was observed when *C. albicans* was treated with 111 µg/mL (Figure 7A). Dose-dependent inhibition activity was observed for all concentrations below 111 µg/mL against both *C. albicans* and *C. neoformans* (Figure 7B). All concentrations below 111 µg/mL showed no effect on the beginning of the exponential phase while the growth steadily increased with no stationary phase up to 12 h. Otherwise, the growth of *C. albicans* steadily increased after 9 h, representing the beginning of the exponential phase, when it was treated with 37 µg/mL.

For *n*-BuOH fraction, the complete inhibition of the visible growth of *C. albicans* and *C. neoformans* was observed at concentrations higher than 111 and 37 µg/mL (Figure 7D and E), respectively. All concentrations below 333 µg/mL and below 111 µg/mL showed dose-dependent inhibition activity against *C. albicans* and *C. neoformans*, respectively. The concentrations of 111 and 37 µg/mL showed remarkable growth inhibition of *C. albicans* but the beginning of the exponential phase was not affected. The concentration of 37 µg/mL led to the delay of the beginning of the exponential phase of *C. neoformans* for 1 hour. All other concentrations below 37 µg/mL showed normal beginning of the exponential phase of *C. neoformans* while no stationary phase was observed.

**Minimum Inhibitory Concentration (MIC) and Minimum bactericidal concentration (MBC) of *N. exigua* fractions**

MIC is defined as the lowest concentration that inhibits the visible growth of the tested microbes after 24 h. The results of larger inhibition zones reflect lower MICs. As compared to the influence on the growth curves, persistent effects of all fractions were observed at 12 h and 24 h incubation. The lowest concentrations that completely inhibited the visible growth of all microbes after 12 h were same even after 24 h.

The lowest MIC (Table 2) observed was 37 µg/mL for *n*-BuOH fraction against *S. aureus*. The MIC of $CH_2Cl_2$ fraction against *S. aureus* was 111 µg/mL. The same MICs (111 µg/mL) were observed for $CH_2Cl_2$ and *n*-BuOH fractions against *B. cereus*. Meanwhile, the lowest MIC indicated against yeasts was 111 µg/mL for both $CH_2Cl_2$ and *n*-BuOH fractions against *C. neoformans*. The MICs against *C. albicans* were also equal for both fractions, which is 333 µg/mL.

The MBC is defined as the lowest concentration required in killing the tested microbe. The MBC was performed by culturing the wells, which showed complete inhibition of the visible growth, onto agar plates. The concentration that showed no colonies (0 CFU) on the agar plate, after incubation for 24 h for bacteria or 48 h for yeasts, was considered as the MBC.

The MBCs (Table 2) of all fractions tested were higher than the MICs. The exception of this was the MBC of $CH_2Cl_2$ fraction against *S. aureus*. The MBC of $CH_2Cl_2$ fraction against *S. aureus* was equal to its MIC, which is 37 µg/mL. Besides, the MBC of *n*-BuOH fraction against *S. aureus* was 333 µg/mL, which is similar to MBCs of $CH_2Cl_2$ and *n*-BuOH fractions against *C. neoformans*. In contrast, all fractions tested against *B. cereus* and *C. albicans* showed no killing activity up to the highest concentration tested (1000 µg/mL).

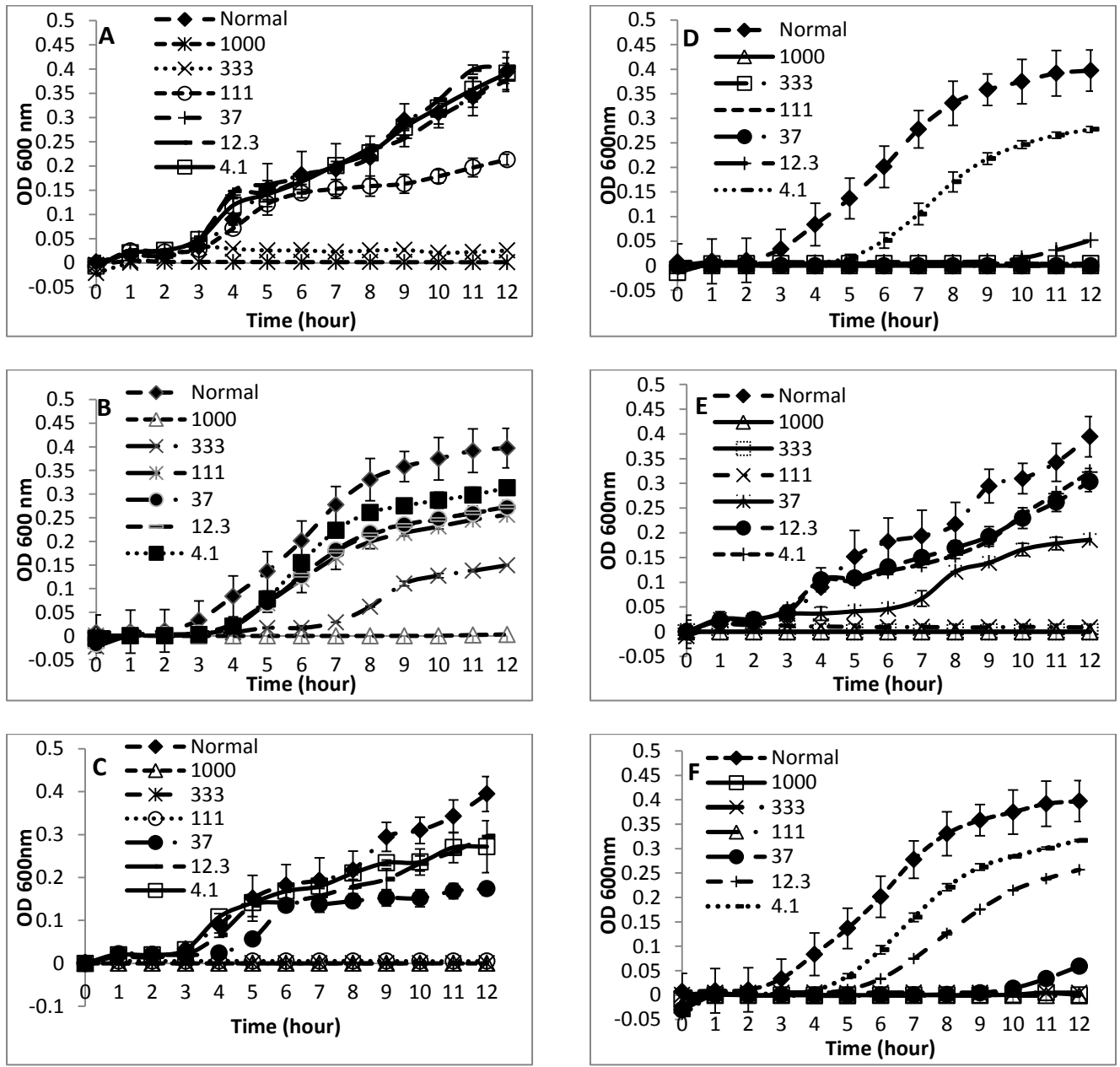

**Figure 6:** Influence of *n*-hexane, CH$_2$Cl$_2$ and *n*-BuOH fractions at low concentrations (1000-4.1 µg/mL) on the growth of Gram-positive bacteria in MHB media during 12 h of incubation at 37 °C. A: *B. cereus* treated with hexane fraction; B: *S. aureus* treated with hexane fraction; C: *B. cereus* treated with CH2Cl2 fraction; D: *S. aureus* treated with CH2Cl2 fraction; E: *B. cereus* treated with n-BuOH fraction; F: *S. aureus* treated with n-BuOH fraction. Data represent mean of triplicate of the growth turbidity and expressed as OD$_{600}$±SD.

**Table 2:** Minimum Inhibitory Concentration (MIC) and Minimum bactericidal concentration (MBC) of *N. exigua* fractions:

| Fraction | *B. cereus* | | *S. aureus* | | *C. albicans* | | *C. neoformans* | |
|---|---|---|---|---|---|---|---|---|
| | MIC | MBC | MIC | MBC | MIC | MBC | MIC | MBC |
| *n*-Hexane | 1000 | >1000 | 1000±0.0 | >1000 | - | - | - | - |
| CH$_2$Cl$_2$ | 111 | >1000 | 37 | 37 | 333 | >1000 | 111 | 333 |
| *n*-BuOH | 111 | >1000 | 111 | 333 | 333 | >1000 | 111 | 333 |
| Positive control | 20 | - | 20 | - | 400 | - | 300 | - |

Data represent mean of triplicates and expressed as µg/mL±SD; -: not determined.
Positive control: Chloramphenicol or Amphotericin B

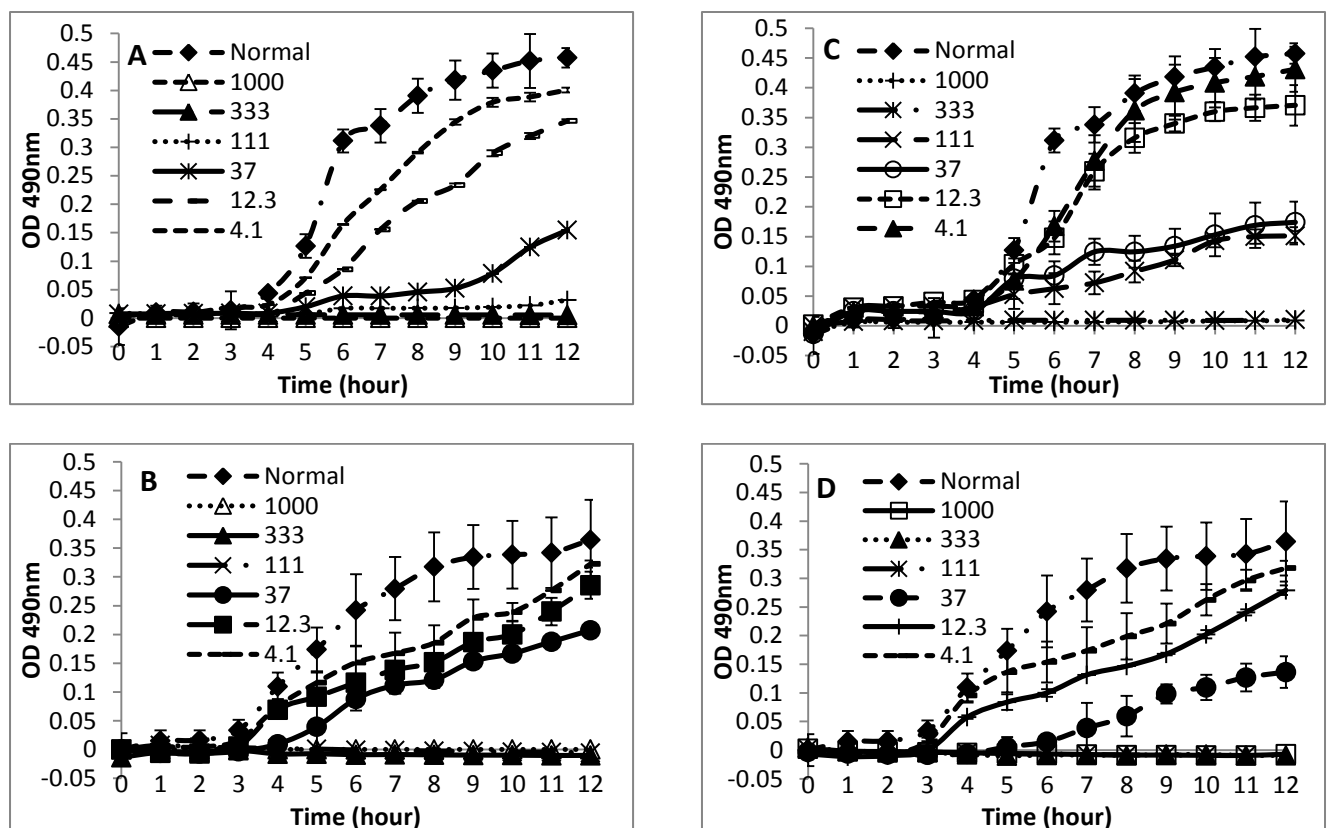

**Figure 7:** Influence of CH$_2$Cl$_2$ and *n*-BuOH fractions at low concentrations (1000-4.1 µg/mL) on the growth of yeast strains in PDB media during 12 h of incubation at room temperature. A: *C. albicans* treated with CH2Cl2 fraction, B: *C. neoformans* treated with CH2Cl2 fraction. C: *C. albicans* treated with n-BuOH fraction, D: *C. neoformans* treated with n-BuOH fraction. Data represent mean of triplicate of the growth turbidity and expressed as OD$_{490}$±SD.

## DISCUSSION

Two methods were used in this study to evaluate the antimicrobial activity. The preliminary screening was done using disc diffusion method while the most accurate potency of the extract was indicated by microdilution method. The use of disc diffusion method is useful in the initial screening for antimicrobial activity (Jenkins et al., 1998). Equally important is the advantage of providing the absolute value or concentration at which an extract is effective against the test microorganism by using microdilution method (Koh et al., 2002). Besides, the absence of antimicrobial activity in disc diffusion method does not necessarily indicate lack of antimicrobial activity. This might be due to the rate of diffusion of the extract into the agar and the potency of the extract (Kelman et al., 2006). In microdilution method, the solubility of the specific antimicrobial constituents of the extracts might be better (Espina et al., 2011). The influence on the growth curves is known as the best method in evaluation of antimicrobial agents. This method provides a descriptive relationship between the tested sample and the microbes (Patton et al., 2006; Rufián-Henares and Morales, 2008).

An attempt was made in this study to investigate the origin of the metabolites that represent the antimicrobial activity of the marine sponge *N. exigua*. *N. exigua*-associated bacteria were isolated and evaluated in two forms; total associated microbes (mixed culture containing all *N. exigua* associates) and *N. exigua* isolates (pure culture). The results showed that the total-associates have no antimicrobial activity while only three marine isolates showed negligible antimicrobial activity against *B. cereus*. On the other hand, *N. exigua* biomass extract exhibited potent antimicrobial activity. In fact, the presence of even one millimetre inhibition zone indicates the presence of active metabolites. Therefore, the preliminary investigation of the antimicrobial properties of *N. exigua* provided valuable hints about the abundance of the active metabolites but not about the origin of the active metabolites. Lin et al., (2001) compared the secondary metabolites of fungus *Aspergillus versicolor*, which were cultured from *N. exigua*, with those of *N. exigua* itself. They found that the secondary metabolites isolated from the sponge were not from the fungus *Aspergillus versicolor* since they were not similar.

Faulkner et al., (1994) hypothesised that the sponge, as filter-feeder, filtrates and concentrates the active metabolites from a microorganism that is abundant in the water. Many compounds have been isolated from different sources that possess the same or similar chemical structures. It is highly possible that the pyridoacridine alkaloids, which are isolated from sponges, tunicates and coelenterate, are produced by microorganisms (Molinski, 1993). Another example is renieramycin-type alkaloids, which are isolated from Xestospongia sponges. The chemical structure of renieramycin is analogous with saframycins and safrins, which are isolated from Streptomyces bacterial metabolites (Davidson, 1992).

In our previous study (Qaralleh et al., 2011), aqueous extract of *N. exigua* showed higher antimicrobial activity than its organic extract. Marine bioactive compounds are extremely polar, thus aqueous solutions or strongly polar solvents such as methanol must be used for extraction (Houssen and Jaspars, 2005; Riguera, 1997). But, the aqueous solutions are usually avoided since they are highly problematic. Problems like bacterial and fungal growth may lead to degradation of the active components. Production of endotoxins by microbes may also give false results in bioassays. Aqueous solutions need high temperature for evaporation; the temperature may cause degradation of the metabolites and may also induce the microbial growth. On the other hand, the presence of salts in aqueous extracts makes the process of isolation of the metabolites in a pure form more difficult (Shimizu, 1985; Shimizu, 1998; Wright, 1998; Houssen and Jaspars, 2005). Hence, *N. exigua* metabolites were extracted in this study using methanol.

In drug discovery programmes, active natural product is extracted from the source, concentrated, fractionated, and purified, yielding one or more pure biologically active compounds (Koehn, 2008). Further pharmacological assays and chemical work should be carried out if the pure compound shows novel bioactivity (Riguera, 1997). In this report, bio-guided isolation to localise the active component of *N. exigua* was conducted according to the recommendation by Riguera (1997). A sequential gradient partition with different solvents was performed to obtain fractions containing compounds distributed according to their polarity. The antimicrobial activities of *N. exigua* fractions were then evaluated using disc diffusion and microdilution methods. The results showed that the active metabolites were present in *n*-hexane, $CH_2Cl_2$, *n*-BuOH, and water fractions. The presence of the inhibition activities in these fractions might indicate the presence of many active metabolites with varied polarities. Such metabolites like alkaloid salts, amino acids, polyhydroxy steroids, and saponins are predicted to be found in the *n*-BuOH fraction. The $CH_2Cl_2$-soluble fraction gain compounds of medium polarity such as peptides and depsipeptides, while in the *n*-hexane and $CCl_4$ fractions, low polarity metabolites like hydrocarbons, fatty acids, acetogenins, terpenes, and alkaloids could be found (Riguera, 1997; Ebada et al., 2008).

The presence of the antimicrobial activity distributed in different polarity fractions, is in agreement with many previous studies. Galeano and Martínez (2007) screened the antimicrobial activity of 24 sponge species extracted with different polarity solvents. The study revealed the presence of active metabolites distributed in methanol, chloroform, and *n*-hexane fractions (Galeano and Martínez, 2007). Similar results were obtained by the study conducted by Safaeian et al., (2009). According to the study, antimicrobial activity was found in different polarity extracts (methanol, ethyl acetate, and *n*-hexane) suggesting the presence of diverse active metabolites.

The antimicrobial activities of the relatively polar fractions ($CH_2Cl_2$ and *n*-BuOH) in this study showed greater potency compared to the *n*-hexane and water fractions. In fact, sponges' potent antimicrobial metabolites are usually found in polar fractions such as $CH_2Cl_2$ and *n*-BuOH. However, some antimicrobial metabolites have been isolated from low polarity fractions such as *n*-hexane and benzene. Two novel anti-tuberculosis parguesterols have been isolated from the *n*-hexane-soluble fraction of the Caribbean sponge, *Svenzea zeai* (Wei et al., 2007). Lakshmi et al., (2010) isolated araguspongin C from the *n*-hexane fraction of *Haliclona exigua*. Araguspongin C possesses promising antifungal activity (Lakshmi et al., 2010). The benzene extract of *Xestospongia exigua* has yielded pentacyclic polyketide, which possesses antimicrobial activity against *S. aureus* and *B. subtilis* (Roll et al., 1983).

In this study, the antimicrobial activity was evaluated using different methods. Generally, dose-dependent antimicrobial activities of *N. exigua* fractions were observed. *N. exigua* $CH_2Cl_2$ and *n*-BuOH fractions were the most potent fractions. The effect of $CH_2Cl_2$ and *n*-BuOH fractions against *S. aureus* and *C. neoformans* was not only bacteriostatic, but it was also bactericidal because it caused death to the microbes. In fact, an agent with *in vitro* bactericidal activity is preferable compared to one with only *in vitro* bacteriostatic activity (Pankey and Sabath, 2004). Since the *n*-hexane fraction showed no bactericidal activity, the $CH_2Cl_2$ and *n*-BuOH fractions can be considered as the optimal fractions to be used in bio-guided isolation assay since they showed potent bactericidal activities.

In all microbes tested, the larger inhibition zone was reflected in lower MIC. But *B. cereus* was more susceptible to *N. exigua* fractions compared to *S. aureus* in the disc diffusion method. In contrast, the influence on the growth curves of *B. cereus* and the MICs revealed different sensitivities to the antimicrobial activity. In addition, there was no killing action for all fractions tested against *B. cereus*. Notably, clear inhibition zones were detected when *N. exigua* fractions were tested against *B. cereus* but a zone of growth was also observed around the disc. The zone of growth was found at concentrations of 1, 2, and 2.5 mg/disc but it was absent at concentration of equal to 0.4 mg/disc. Therefore, the ability of *B. cereus* to produce endospores under harsh conditions (Saz, 1970; Fenselau et al., 2008) was assumed to be the reason. To confirm this, endospore staining was performed for the bacteria obtained from the zone of growth (Figure 8, pointer A) and bacteria far from the zone of inhibition in the same plate (Figure 8, pointer B). As a result, high percentage of the bacteria from the zone of growth appeared to be in endospore form compared to the bacteria further away from the same plate. The presence of endospores was also reflected on the MBCs of *N. exigua* fractions against *B. cereus*. The MBCs of all fractions tested against *B. cereus* were more than 1000 µg/mL.

In agar-based methods like disc diffusion method, the metabolites uploaded into the disc diffuse into the agar creating a concentration gradient. The concentrations of the metabolites decrease when the distance from the disc increases. Therefore, the area near the disc contains the highest concentration of the metabolites (Kiska, 1998; Othman et al., 2011). This may explain that the highest concentrations tested (1, 2, and 2.5 mg/disc) induced sporulation in *B. cereus* around the disc compared to the lowest concentration (0.4 mg/disc).

The difference between bactericidal and bacteriostatic agents is clear according to the *in vitro* definition (Pankey and Sabath, 2004). The pattern of activity, indicated by the influence of *N. exigua* fractions on the growth curves, suggests that the fractions tested at certain concentrations are bacteriostatic against all microbes.

## CONCLUSION
In conclusion, the abundance of the antimicrobial metabolites appears to be in the sponge biomass. *N. exigua* biomass extract showed potent antimicrobial activity. The potential of antimicrobial activity of *N. exigua* fractions revealed the presence of more than one active metabolite. *S. aureus* was the most susceptible microbe evaluated. *n*-hexane, $CH_2Cl_2$, and *n*-BuOH fractions have been selected to be subjected to bio-guided isolation using different chromatographic techniques in order to isolate the active metabolites.

## ACKNOWLEDGMENT
This work was supported by a grant from Research Management Center, International Islamic University Malaysia, IIUM (EDW B 10-107-0446). The authors would like to thank the staff of INOCEM (Institute of Oceanography and Maritime Studies, IIUM) for their hospitality and kind assistance in field collection. Acknowledgements to Mr. Lim Swee Cheng (Tropical Marine science institute, National university of Singapore, Singapore) for assistance during sponge identification.